\documentclass[fleqn,twoside,twocolumn,nofootinbib,showkeys]{revtex4} 
\usepackage[sec,nocpr]{ujp} 
\begin{document}
\title[Structure of \boldmath$^{14}$C and \boldmath$^{14}$O Nuclei]
{STRUCTURE OF \boldmath$^{14}$C AND \boldmath$^{14}$O NUCLEI\\
CALCULATED IN THE VARIATIONAL
APPROACH}%

\author{B.E.~Grinyuk}
\affiliation{\bitp}
\address{\bitpaddr}
\email{bgrinyuk@bitp.kiev.ua, \, dvpyat@gmail.com}

\author{D.V.~Piatnytskyi}
\affiliation{\bitp}
\address{\bitpaddr}

\udk{539.172} \pacs{27.20.+n,  21.60.Gx,\\[-3pt]  21.10.Ft,
21.10.Gv} \razd{\secii}

\autorcol{B.E.\hspace*{0.7mm}Grinyuk,
D.V.\hspace*{0.7mm}Piatnytskyi}

\setcounter{page}{674}%

\begin{abstract}
The structure of mirror $^{14}$\!C and $^{14}$\!O nuclei has been
studied in the framework of the five-particle model (three
$\alpha$-particles and two nucleons).\,\,Interaction potentials are
proposed, which allowed the energy and radius of $^{14}$\!C nucleus,
as well as the energy of $^{14}$\!O one, to agree with experimental
data.\,\,On the basis of the variational approach with the use of
Gaussian bases, the energies and wave functions for five-particle
systems under consideration are calculated.\,\,The charge radius of
$^{14}$\!O nucleus, as well as the charge density distributions and
the form factors for both nuclei, are predicted.
\end{abstract}


\maketitle

\section{Introduction}

Each of the radioactive $^{14}$C and $^{14}$O nuclei can be imagined
as composed of three $\alpha$-particles and two extra
nucleons.\,\,The experience of the theoretical researches of nuclei
with two extra nucleons such as $^{6}$He and $^{6}$Li
\cite{R1,R2,R3} or more complicated $^{10}$Be and $^{10}$C ones
\cite{R4,R5} showed that the accuracy of this approximation can
compete with that of other approaches, in which all nucleon degrees
of freedom are taken into consideration \cite{R6}.\,\,If the number
of nucleons in a nucleus is large, the reduction of the number of
variables in the corresponding problem by making allowance for the
$\alpha$-clustering becomes even more \mbox{justified.}\looseness=1

In this work, we consider $^{14}$C and $^{14}$O nuclei as systems
consisting of three $\alpha$-particles and two extra neutrons (in
$^{14}$C case) or two extra protons (for $^{14}$O nucleus).\,\,The
problem of five particles is solved in the variational approach with
the use of the Gaussian basis \cite{R7,R8,R9}, which allows the
systems of several particles with various kinds of interaction
between them to be studied with a rather high accuracy.

The applied models of generalized interaction potentials between
$\alpha $-particles, as well as between nucleons and
$\alpha$-particles, are  similar to those used by us in the research
of lighter nuclei \cite{R3,R5,R10}.\,\,Ho\-we\-ver, for the
agreement of the energy and charge radius of $^{14}$C nucleus and
the energy of $^{14}$O nucleus with experimental values to be exact,
the parameters of potentials are slightly modified.\,\,This gives us
a ground to hope for that the density distributions and charge form
factors predicted for $^{14}$C and $^{14}$O nuclei~-- in particular,
the charge radius of $^{14}$O~-- will agree with future experimental
data.\vspace*{-2mm}

\section{Statement of the Problem}

The model Hamiltonian for $^{14}$O nucleus contains, besides the
one-particle operator of kinetic energy, the potentials of pair
interaction between the particles, which are generated by nuclear
forces, and the Coulomb repulsion potential,\vspace*{-2mm}
\[
\hat{H}=\sum\limits_{i=1}^{2}\frac{\mathbf{p}_{i}^{2}}{2m_{p}}+\sum
\limits_{i=3}^{5}\frac{\mathbf{p}_{i}^{2}}{2m_{\alpha}}+U_{pp}\left(
r_{12}\right) +
\!\!\sum\limits_{j>i=3}^{5}\!\!\hat{U}_{\alpha\alpha}\left(
r_{ij}\right) +
\]\vspace*{-7mm}%
\begin{equation}
+\sum\limits_{i=1}^{2}\sum\limits_{j=3}^{5}\hat{U}_{p\alpha}\left(
r_{ij}\right)  +\sum\limits_{j>i=1}^{5}\frac{Z_{i}Z_{j}e^{2}}{r_{ij}}.
\label{E1}%
\end{equation}
Here, $m_{p}$ and $m_{\alpha}$ are the masses of a proton and an
$\alpha $-particle, respectively;  $Z_{1}=Z_{2}=1$ and
$Z_{3}=Z_{4}=Z_{5}=2$ are the charges of particles divided by the
elementary charge unit $e$.\,\,Note that the Hamiltonian for
$^{14}$C nucleus differs from Eq.~(\ref{E1}) in that the neutron
charges equal zero ($Z_{1}=Z_{2}=0$) and the proton mass $m_{p}$ is
substituted by the neutron one $m_{n}$, although this change of mass
practically does not affect the result.\,\,The Hamiltonians for both
nuclei are also characterized by a small difference between the
effective nuclear interaction of a neutron with an $\alpha$-particle
in $^{14}$C nucleus and the effective interaction of a proton with
an $\alpha$-particle in $^{14}$O one, because the distribution of
protons in the $\alpha$-particle has a little larger radius than the
corresponding distribution of neutrons (see, e.g.,
work~\cite{R11}).

The choice of models for the pair interaction potentials between the
particles is based on the criterion that our model should
simultaneously describe the experimental energies of the examined
nuclei and the charge radius of $^{14}$C nucleus (the experimental
charge radius for $^{14}$O is unknown).\,\,The pair interaction
potential between the neutrons in $^{14}$C nucleus is chosen in the
form of a local potential that describes both the low-energy singlet
neutron-neutron parameters and the singlet phase of the neutron
scattering with a qualitative accuracy.\,\,In view of the charge
invariance of nuclear forces, the interaction potential between the
protons in $^{14}$O nucleus, $U_{pp}\left(  r\right)  $, is assumed
to be the same.\,\,This potential was successfully used, while
studying $^{6}$He \cite{R3}, $^{10}$Be, and $^{10}$C \cite{R5,R10}
\mbox{nuclei.}

The interaction potential between a neutron and an $\alpha$-particle
is chosen in the form of a generalized interaction potential with
the local and nonlocal components, which models most successfully
the interaction of a nucleon with the $\alpha$-cluster \cite{R1,R3}
and simulates the Pauli exclusion principle.\,\,In work \cite{R5},
in order to study $^{10}$Be nucleus, the parameters of the potential
between a neutron and an $\alpha$-particle that had been used to
study $^{6}$He nucleus \cite{R3} were slightly changed.\,\,In the
present work, analogously to this procedure, some parameters of the
same potential are changed a little in order to simultaneously
describe the energy and charge radius of $^{14}$C with a high
accuracy:
\begin{equation}
\hat{U}_{n\alpha}\left(  r\right)  =-V_{0}\exp(  -( r/r_{0})
^{2})  +\frac{g}{\pi^{3/2}R_{0}^{3}}|u\rangle\langle u|, \label{E2}%
\end{equation}
where the local attraction is characterized by the parameters $V_{0}%
=43.95\mathrm{~MeV}$ and $r_{0}=2.25$~fm, and the separable
repulsion has the form factor $u(r)=$ $=\exp\left(  -\left(
r/R_{0}\right)  ^{2}\right)  $ with the radius
$R_{0}=2.79~\mathrm{fm}$ and the repulsion intensity $g=140\lambda
\left(  R_{0}\right)  $~MeV (hereafter, $\lambda\left(  x\right)
\equiv\left[  2/(\pi x^{2})\right]  ^{3/2}$).

The interaction potential between a proton and an $\alpha$-particle
that was used by us to study $^{14}$O nucleus had the same form
(\ref{E2}), but with slightly changed parameters.\,\,This
circumstance is related to the known fact that the protons and
neutrons in $^{4}$He nucleus are not distributed absolutely
identically (the details of distributions can be found in work
\cite{R11}), so that the potential of nuclear interaction $\hat
{U}_{n\alpha}$ should not exactly coincide with the potential $\hat
{U}_{p\alpha}$.\,\,We select the parameters for the potential $\hat
{U}_{p\alpha}$ so that the energy of $^{14}$O nucleus could be
described by assuming the other potentials of nuclear interaction
between the corresponding particles in $^{14}$O and $^{14}$C nuclei
to be identical.\,\,In this work, the attraction intensity
$V_{0}=44.757~$MeV and the attraction radius $r_{0}=2.27~$fm are
used for the potential $\hat{U}_{p\alpha}$.\,\,The separable
repulsion component of this potential is taken the same as for the
potential $\hat{U}_{n\alpha}$~(\ref{E2}).\looseness=1

By its form, the interaction potential between $\alpha$-particles is
also very similar to that used in work \cite{R5}:\,\,it has local
and nonlocal components, but with slightly modified parameters:
\[
\hat{U}\left(  r\right)  =-U_{1}\exp\left(  -\left(  r/\rho_{1}\right)
^{2}\right)  +
\]\vspace*{-5mm}
\begin{equation}
+\,U_{2}\exp\left(  -\left(  r/\rho_{2}\right)  ^{2}\right)
+\frac{g}{\pi
^{3/2}\rho_{\alpha}^{3}}|v\rangle\langle v|, \label{E3}%
\end{equation}
where the attraction intensity $U_{1}=43.5~$MeV, the repulsion intensity
$U_{2}=240.0~$MeV, and the corresponding radii equal $\rho_{1}=2.55$%
~\textrm{fm} and $\rho_{2}=$ $=1.3~\mathrm{fm}$,
respectively.\,\,The separable repulsion has the form factor
$v(r)=\exp\left(  -\left( r/\rho_{\alpha }\right)  ^{2}\right)  $
with the radius $\rho_{\alpha}=1.765~$fm, and the repulsion
intensity $g=60\lambda\left(  \rho_{\alpha}\right)~$MeV.

Note that the selection of parameters for the potentials became
possible only after a multiply repeated procedure that includes the
solution of the five-particle problem and the comparison of the
values obtained for the energy and the charge radius with the
experimental data.\,\,The method of calculation on the basis of the
variational approach with the use of Gaussian bases will be briefly
described in the next section.\,\,The resulting potential parameters
are given above, and the energies and radii calculated on the basis
of those potentials for both analyzed nuclei are compared with the
corresponding experimental data in Table~1.\,\,In the latter, the
energies of nuclei are shown subtracting $-28.296$~MeV per each
$\alpha$-particle, and the charge radii were calculated with regard
for the non-point character of the particles in the Helm
approximation:
\[
R_{\mathrm{ch}}^{2}=R_{\alpha}^{2}+R_{\mathrm{ch}}^{2}\left(  ^{4}%
\mathrm{He}\right)
\]
in the case of $^{14}$C nucleus (a small squared charge radius of
a neutron is neglected) and
\[
R_{\mathrm{ch}}^{2}=\frac{3}{4}\left(  R_{\alpha}^{2}+R_{\mathrm{ch}}%
^{2}\left(  ^{4}\mathrm{He}\right)  \right)  +\frac{1}{4}\left(  R_{p}%
^{2}+R_{\mathrm{ch}}^{2}\left(  p\right)  \right)
\]
for $^{14}$O one.\,\,Here, $R_{\alpha}$ designates the
root-mean-square radius for \textquotedblleft point-like\textquotedblright%
\ $\alpha$-particles distribution, which was calculated in the
framework of the model with Hamiltonian (\ref{E1}), and the charge
radius of $\alpha$-particle $R_{\mathrm{ch}}\left(
^{4}\mathrm{He}\right) =1.679~$fm was taken from the experiment (as
an average value between the modern data \cite{R12} and the data of
work \cite{R13}).\,\,Analogously, $R_{p}$ means the calculated
root-mean-square radius of the \textquotedblleft point-like\textquotedblright%
\ proton distribution (in the case of $^{14}$O nucleus), and $R_{\mathrm{ch}%
}\left(  p\right)  =0.875~$fm was taken from the experimental data of work
\cite{R14}.

\begin{table}[b]
\noindent\caption{Energies (MeV) and charge radii (fm)\\ of
\boldmath$^{14}$C and $^{14}$O nuclei. The energies are reckoned\\
from the threshold of the nucleus decay into three\\ $\alpha$-particles
and two nucleons} \vskip3mm\tabcolsep8.6pt

\noindent{\footnotesize\begin{tabular}{|c| c| c| c| c| c| c| }
 \hline \multicolumn{1}{|c}
{\rule{0pt}{5mm}Nucleus} & \multicolumn{1}{|c}{$E$} &
\multicolumn{1}{|c}{$E_{\exp}$}& \multicolumn{1}{|c}{$R_
{\mathrm{ch}}$} & \multicolumn{1}{|c|}{$R_
{\mathrm{ch},\,\mathrm{exp}}$}
\\[2mm]%
\hline%
 \rule{0pt}{5mm}$^{14}$C
 & $-20.398$ & $-20.398$ & $2.500$ & $2.496$ \cite{R17}
\\[-0.5mm]
 &  &  &  & $2.503$ \cite{R12}
\\[1.5mm]

$^{14}$O & $-13.845$ & $-13.845$ & $2.415$ &
---
\\[2mm]
 \hline
\end{tabular}}
\end{table}

\section{Calculation Technique}

To solve the problem of bound states in the system of five
particles, we use the variational method in the Gaussian
representation \cite{R7,R8,R9}, which allows the wave function of
the system to be obtained in the explicit convenient form of a
gaussoid superposition.\,\,Omitting the details of this well-known
method, we only recall that the Schr\"{o}dinger equation with
Hamiltonian (\ref{E1}) can be reduced to a system of linear
algebraic equations (the Galerkin method):
\begin{equation}
\sum_{m=1}^{K}C_{m}\left\langle\! \hat{S}\varphi_{k}\left\vert \hat
{H}-E\right\vert \hat{S}\varphi_{m}\!\right\rangle =0,\quad
k\,=\,0,1,...,K.\label{E4}%
\end{equation}
Here, all required matrix elements are expressible in an explicit
form if one uses the Gaussian basis.\,\,For the ground symmetric
$J^{\pi}=0^{+}$ state, the wave function $\Phi$ has a simple form in
the Gaussian repre\-sentation:
\[
\Phi=\hat{S}\sum_{k=1}^{K}C_{k}\varphi_{k}\equiv
\]\vspace*{-5mm}
\begin{equation}
\equiv\hat{S}\sum_{k=1}^{K}C_{k}\,\exp\!\left(\!  -\sum_{j>i=1}^{5}%
a_{k,ij}\left(  \mathbf{r}_{i}-\mathbf{r}_{j}\right)  ^{2}\!\right)\!\!,\label{E5}%
\end{equation}
where $\hat{S}$ is the symmetrizing operator.\,\,Note also that the
wave function symmetrization can be either performed or not
performed explicitly, because, as was shown in work \cite{R15}, the
symmetric form is restored automatically, when the non-symmetrized
basis is expanded.\,\,Both opportunities are used in this
work.\,\,The explicit symmetrizations for three $\alpha $-particles
and two extra nucleons (in this case, there are only
$3!\times2=12$~terms in the symmetrized function) allowed a Gaussian
basis with considerably smaller dimensionality to be used than that
required in the case of the wave function with no explicit
symmetrization, provided the same calculation
\mbox{accuracy.}\looseness=1

\section{Density Distributions\\ and Form Factors}

The one-particle density distribution for the $j$-th particle in a
system of particles with the wave function $\left\vert
\Phi\right\rangle $ is defined as follows:
\begin{equation}
n_{i}\left(  r\right)  =\left\langle \Phi\right\vert \delta\left(
\mathbf{r}-\left(  \mathbf{r}_{i}-\mathbf{R}_{\mathrm{c.m.}}\right)  \right)
\left\vert \Phi\right\rangle\!, \label{E6}%
\end{equation}
where $\mathbf{R}_{\rm c.m.}$ is the radius vector of the center of
mass of the system.\,\,Hereafter, all density distributions are
normalized to 1: $\int n_{i}\left(  r\right)  d\mathbf{r}=1$.\,\,The
expressions for $n_{i}\left(  r\right) $ have the explicit  form in
terms of the parameters $a_{k,ij}$ and the coefficients $C_{k}$ of
the linear expansion of the wave function in the Gaussian
representation (\ref{E5}).\,\,The resulting one-particle density
distributions multiplied by $r^{2}$ are shown in Fig.~1 by dashed
curves.\,\,They illustrate the density distributions for
\textquotedblleft point-like\textquotedblright\ $\alpha$-particles
and extra nucleons in $^{14}$C nucleus.\,\,The corresponding
distributions of \textquotedblleft point-like\textquotedblright\
particles in $^{14}$O nucleus are very similar to those depicted in
Fig.\,\,1, so that they are not exhibited separately.\,\,The
attention is attracted by the fact that the extra nucleons are
mainly located inside $^{12}$C cluster formed by $\alpha$-particles
(with a probability of about 0.86 for neutrons in $^{14}$C nucleus
and 0.84 for protons in $^{14}$O nuc\-leus).\,\,Ho\-we\-ver, another
small maximum in the density distribution curve for extra nucleons
testifies that the extra nucleons can also be found outside $^{12}$C
cluster, although with rather a low probability (approximately 0.14
for $^{14}$C and 0.16 for~$^{14}$O).\looseness=1

Note that the extra nucleons move much more rapidly than the $\alpha
$-particles.\,\,In particular, the calculated average kinetic energy
amounts to about 32.66~MeV for each of the extra neutrons in
$^{14}$C nucleus.\,\,At the same time, the corresponding value for
each $\alpha$-particle equals about 6.83~MeV, which is almost five
times lower and can be explained, mainly, by the larger mass of the
latter.\,\,Concerning the particle velocities, it turns out that the
extra neutrons in $^{14}$C nucleus move approximately 4.4 times
faster than the $\alpha$-particles.\,\,The same ratio between the
velocities of extra nucleons and $\alpha$-particles is
characteristic of $^{14}$O nucleus.\,\,For the latter, the
calculated average kinetic energy amounts to about 31.77~MeV for
each extra proton and to about 6.62~MeV for each $\alpha$-particle.

To calculate the charge density distribution in the nuclei, the
non-point nature of $\alpha$-particles, as well as protons in the
case of $^{14}$O nucleus, has to be taken into consideration.\,\,For
this purpose, we use the Helm approximation \cite{R16}.\,\,In
particular, the charge density distribution for $^{14}$C nucleus,
\begin{equation}
n_{\mathrm{ch}}\left(  r\right)  =\int n_{\alpha}\left(  \left\vert
\mathbf{r}-\mathbf{r}^{\prime}\right\vert \right)  n_{\mathrm{ch,^{4}He}%
}\left(  r^{\prime}\right)  d\mathbf{r}^{\prime}\!, \label{E7}%
\end{equation}

\begin{figure}%
\vskip1mm
\includegraphics[width=\column]{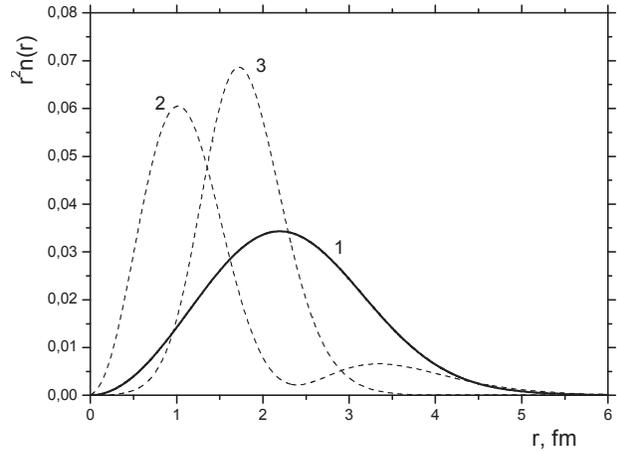}
\vskip-3mm\caption{Charge density distribution in $^{14}$C nucleus
multiplied by $r^{2}$ (solid curve~\textit{1}). Dashed curves
correspond to the density distributions (multiplied by $r^{2}$) of
\textquotedblleft point-like\textquotedblright\ particles: extra
neutrons (curve~\textit{2}) and $\alpha $-particles (curve
\textit{3})}\vspace*{2mm}
\end{figure}

\noindent looks like the convolution of the one-particle density
distribution
$n_{\alpha}$ found for \textquotedblleft point-like\textquotedblright%
\ $\alpha$-particles with the charge density of an $\alpha$-particle
itself, $n_{\mathrm{ch,^{4}He}}$, which follows from the
experimental form factor \cite{R18}.\,\, In expression (7), we
neglect a small contribution of extra neutrons to the charge density
distribution.\,\,Note that the Helm approximation (\ref{E7}) is
obtained, by supposing that the wave function of a nucleus is an
(approximate) product of the wave function obtained for the
Hamiltonian that describes the relative motion of \textquotedblleft
point-like\textquotedblright\ particles with the wave functions of
$^{4}$He nuclei ($\alpha$-clusters).

A similar expression is obtained for $^{14}$O nucleus, in which,
besides the contribution of $\alpha$-particles to the density
distribution, the contribution of extra protons is also made
allowance for:
\[
n_{\mathrm{ch}}\left(  r\right)  =\frac{3}{4}\int n_{\alpha}\left(  \left\vert
\mathbf{r}-\mathbf{r}^{\prime}\right\vert \right)  n_{\mathrm{ch,^{4}He}%
}\left(  r^{\prime}\right)  d\mathbf{r}^{\prime}\,+
\]\vspace*{-5mm}
\begin{equation}
+\,\frac{1}{4}\int n_{p}\left(  \left\vert
\mathbf{r}-\mathbf{r}^{\prime }\right\vert \right)
n_{\mathrm{ch},p}\left(  r^{\prime}\right)
d\mathbf{r}^{\prime}. \label{E8}%
\end{equation}
Here, the charge distribution for a proton, $n_{\mathrm{ch},p}$, is
taken from work \cite{R19}.\,\,The coefficients 3/4 and 1/4 (their
sum equals 1) are proportional to the total charges of
$\alpha$-particles and extra protons, respectively, in $^{14}$O
nucleus.

\begin{figure}%
\vskip1mm
\includegraphics[width=\column]{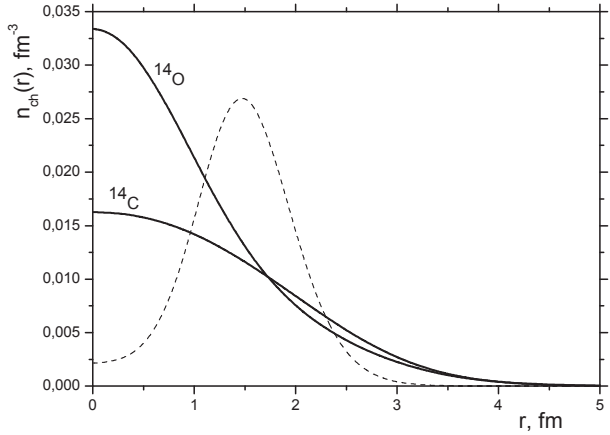}
\vskip-3mm\caption{Charge density distributions in $^{14}$C and
$^{14}$O nuclei. The dashed curve demonstrates the density
distribution for \textquotedblleft point-like\textquotedblright\
$\alpha$-particles in $^{14}$C nucleus. The distributions are
normalized to 1}
\end{figure}

\begin{figure}%
\vskip3mm
\includegraphics[width=7.5cm]{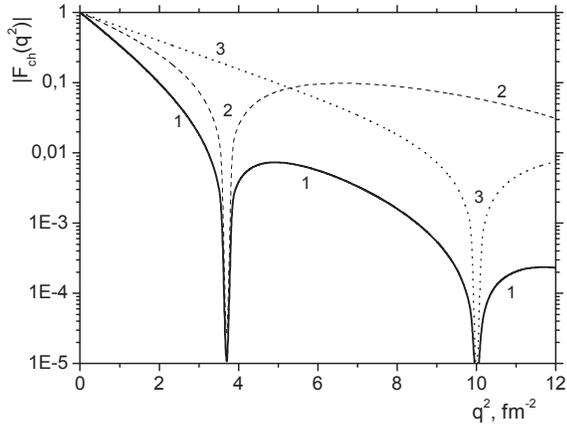}
\vskip-3mm\caption{Charge form factor for $^{14}$C nucleus (solid
curve~\textit{1}) in comparison with the same quantity calculated in
the \textquotedblleft point-like\textquotedblright-particle
approximation (dashed curve~\textit{2}). Curve~\textit{3}
demonstrates the form factor for {$^{4}$He }nucleus {\cite{R12}}}
\end{figure}

The distributions of charge density in $^{14}$C and $^{14}$O nuclei
obtained on the basis of expressions (\ref{E7}) and (\ref{E8}) are shown in
Fig.~2.\,\,A considerably higher charge density at short distances in $^{14}%
$O nucleus is explained by the presence of additional proton charges
located at rather small distances in this nucleus.\,\,For this
reason, the charge radius of $^{14}$O nucleus turns out to be
smaller than that of $^{14}$C one:
\[
R_{\mathrm{ch}}^{2}\left(  \mathrm{^{14}O}\right)  =\int r^{2}%
n_{\mathrm{ch,^{14}O}}\left(  r\right)  d\mathbf{r}<
\]\vspace*{-7mm}
\begin{equation}
<R_{\mathrm{ch}}^{2}\left(  \mathrm{^{14}C}\right)  =\int r^{2}%
n_{\mathrm{ch,^{14}C}}\left(  r\right)  d\mathbf{r}, \label{E9}%
\end{equation}

\begin{table}[b]
\noindent\caption{Root-mean-square relative\\ distances and radii
(fm) of \boldmath$^{14}$C and $^{14}$O nuclei}
\vskip3mm\tabcolsep6.1pt

\noindent{\footnotesize\begin{tabular}{|c| c| c| c| c| c| c| c| }
 \hline \multicolumn{1}{|c}
{\rule{0pt}{5mm}Nucleus} & \multicolumn{1}{|c}{$r_{NN}$} &
\multicolumn{1}{|c}{$r_{N\alpha}$} &
\multicolumn{1}{|c}{$r_{\alpha\alpha}$}&
\multicolumn{1}{|c}{$R_{N}$}& \multicolumn{1}{|c}{$R_ {\alpha}$} &
\multicolumn{1}{|c|}{$R_ {\mathrm{ch}}$}
\\[2mm]
\hline%
\rule{0pt}{5mm}$^{14}$C
 & $2.621$ & $2.667$ & $3.189$ & $1.786$ & $1.852$ & $2.500$
\\
$^{14}$O & $2.732$ & $2.750$ & $3.239$ & $1.864$ & $1.882$ & $2.415$
\\[2mm]
 \hline
\end{tabular}}
\end{table}

\noindent
although all distributions for \textquotedblleft point-like\textquotedblright%
\ particles in $^{14}$O nucleus have, on the contrary, larger radii
in comparison with their counterparts in $^{14}$C.\,\,This
circumstance and a lower binding energy in $^{14}$O nucleus are
explained mainly by the additional Coulomb repulsion due to extra
protons. \,\,In order to confirm this fact,
 we quote the calculated root-mean-square relative distances between
extra nucleons, $r_{NN}$, nucleon and $\alpha$-particle, $r_{N\alpha}$, and
$\alpha$-particles, $r_{\alpha\alpha}$, as well as the root-mean-square radii
for the distributions of \textquotedblleft point-like\textquotedblright%
\ nucleons, $R_{N}$, and \textquotedblleft point-like\textquotedblright%
\ $\alpha$-particles, $R_{\alpha}$, in both analyzed nuclei in
Table~2.\,\,We hope for that the value of charge radius $R_{\rm
ch}\left( ^{14}\mathrm{O}\right) =2.415~\mathrm{fm}$ predicted by us
for $^{14}$O nucleus will be experimentally confirmed in the future.
With regard for the experimental error for the $\alpha$-particle
radius \cite{R12,R13} and the restricted accuracy of our model, in
which the Helm approximation was used, we evaluate the calculation
error for the charge radius of $^{14}$O nucleus to equal about
$\pm$$0.005~$fm.

\begin{figure}%
\vskip1mm
\includegraphics[width=7.7cm]{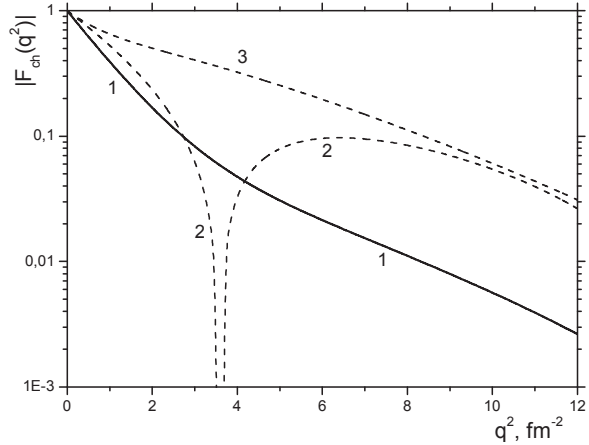}
\vskip-3mm\caption{Charge form factor for $^{14}$O nucleus (solid
curve~\textit{1}), form factor corresponding to the density
distribution of $\alpha$-particles in this nucleus in the
\textquotedblleft point-like\textquotedblright-particle
approximation (curve~\textit{2}), and analogous form factor obtained
for extra protons (curve~\textit{3})}\vspace*{2mm}
\end{figure}

Note that the integration (convolution) in Eqs.~(\ref{E7}) and
(\ref{E8}) substantially \textquotedblleft
smoothes\textquotedblright\ out the one-particle distributions
obtained for \textquotedblleft point-like\textquotedblright\
particles.\,\,As a result, the charge distribution
at short distances does not contain a \textquotedblleft dip\textquotedblright%
\ typical of distributions for \textquotedblleft point-like\textquotedblright%
\ $\alpha$-particles in both nuclei.

Characteristic features in the behavior of density distributions
manifest themselves in the corresponding form factors, the Fourier
transforms of the density.\,\,In particular, convolution (\ref{E7})
transforms into the product
\begin{equation}
F_{\mathrm{ch,^{14}C}}\left(  q\right)  =F_{\mathrm{\alpha,^{14}C}}\left(
q\right)  F_{\mathrm{ch,^{4}He}}\left(  q\right)\!,  \label{E10}%
\end{equation}
and expression (\ref{E8}) into the sum of products
\[
F_{\mathrm{ch,^{14}O}}\left(  q\right)  =\frac{3}{4}F_{\mathrm{\alpha,^{14}O}%
}\left(  q\right)  F_{\mathrm{ch,^{4}He}}\left(  q\right)\,  +
\]\vspace*{-7mm}
\begin{equation}
+\,\frac{1}{4}F_{p\mathrm{,^{14}O}}\left(  q\right)
F_{\mathrm{ch,}p}\left(
q\right)\!  . \label{E11}%
\end{equation}
In the products in expressions (\ref{E10}) and (\ref{E11}), the
first multipliers correspond to the form factors obtained by us from
the wave function of the five-particle problem in the
\textquotedblleft point-like\textquotedblright-particle
approximation.\,\,The second multipliers
are the experimental form factors of an $\alpha$-particle $F_{\mathrm{ch,^{4}He}%
}\left(  q\right)  $ \cite{R18} or proton $F_{\mathrm{ch,}p}\left(
q\right) $ \cite{R19}.\,\,If a form factor becomes zero at a
definite transferred momentum squared, the absolute value of the
form factor has a ``dip'' at this $q^{2}$.\,\,Owing to the
representation of the charge form factor for $^{14}$C nucleus in the
form of product (\ref{E10}), the form factor absolute value has
\textquotedblleft dips\textquotedblright\ at those $q^{2}$, where
each of the multipliers has its own \textquotedblleft
dip\textquotedblright, which is
illustrated in Fig.~3.\,\,Note that the \textquotedblleft dip\textquotedblright%
\ located at the squared transferred momentum
$q^{2}\simeq10$~fm$^{-2}$ and originating from the form factor of an
$\alpha$-particle is also observed in the form factors of $^{6}$He
\cite{R3} and $^{10}$Be \cite{R5} nuclei, as well as in all other
cases where the charge distribution in cluster nuclei is driven only
by the charges of $\alpha$-particles.\,\,Concerning the first dip in
the charge form factor of $^{14}$C nucleus in the interval slightly
below $q^{2}\sim4$~fm$^{-2}$, which is connected with the
\textquotedblleft dip\textquotedblright\ in the first multiplier in
Eq.~(\ref{E10}), it is explained by a substantial decrease in the
density distribution at short distances, which was obtained in the
approximation of \textquotedblleft point-like\textquotedblright\
$\alpha$-particles (in Fig.~2, this distribution is depicted by a
dashed curve).\,\,It can be shown that, after the Fourier
transformation, this distribution transforms into a form factor that
changes its sign in the momentum representation, and the absolute
value of the form factor has a \textquotedblleft
dip\textquotedblright.

At the same time, in the case of $^{14}$O nucleus, owing to the second term
in Eq.~(\ref{E11}), the corresponding \textquotedblleft dips\textquotedblright%
\ do not manifest themselves \textit{per se} in the charge form
factor of the nucleus, but only slightly affect the change of
regimes in its behavior, as is shown in Fig.~4.\,\,Hence, the form
factors of $^{14}$O and $^{14}$C nuclei considerably differ from
each other owing to the role of extra protons in $^{14}$O instead of
neutrons in~$^{14}$C.

\section{Conclusions}

In the framework of the five-particle model (three
$\alpha$-particles and two extra nucleons) and on the basis of
variational calculations in the Gaussian representation, the density
distributions and the form factors for $^{14}$C and $^{14}$O nuclei
are calculated.\,\,The extra nucleons in both nuclei are found to
move predominantly inside $^{12}$C cluster, although they can also
be found at its periphery with a low probability.\,\,The charge
radius of $^{14}$O nucleus is predicted.\,\,It is shown that, owing
to the extra protons that are located closer to the nucleus center,
this radius is smaller than that of $^{14}$C nucleus, despite that
all relative distances between the corresponding particles in
$^{14}$O nucleus are larger than in $^{14}$C one.\,\,We hope for
that the more detailed ideas of the structure of $^{14}$C and
$^{14}$O nuclei and the character of motion of their component
particles can be obtained on the basis of calculations and the
analysis of such structural functions as the pair correlation
functions and the momentum distributions.

\rezume{%
Б.Є.\,Гринюк, Д.В.\,П'ятницький}{СТРУКТУРА ЯДЕР $^{14}$C ТА $^{14}$O
\\ У ВАРІАЦІЙНОМУ ПІДХОДІ} {В рамках п'ятичастинкової моделі (три
$\alpha$-частинки і два додаткові нуклони) досліджено структуру
дзеркальних ядер $^{14}$C та $^{14}$O. Запропоновано потенціали
взаємодії, які дозволили узгодити з експериментом енергію та радіус
ядра $^{14}$C, а також енергію ядра $^{14}$O. На основі варіаційного
підходу з використанням гаусоїдних базисів розраховано енергії і
хвильові функції досліджуваних п'ятичастинкових систем. Передбачено
зарядовий радіус ядра $^{14}$O, а також зарядові розподіли густини і
формфактори обох ядер.}


\begin{thebibliography}{99}                                                                                               %


\bibitem {R1}V.I.~Kukulin, V.N.~Pomerantsev, Kh.D.~Razikov \textit{et al.},
Nucl. Phys. \textbf{A 586}, 151 (1995).\vspace*{0.5mm}

\bibitem {R2}M.V.~Zhukov, B.V.~Danilin, D.V.~Fedorov \textit{et al.}, Phys.
Rep. \textbf{231}, 151 (1993).\vspace*{0.5mm}

\bibitem {R3}B.E.~Grinyuk and I.V.~Simenog, Yad. Fiz. \textbf{72},10 (2009).\vspace*{0.5mm}

\bibitem {R4}Y.~Ogawa, K.~Arai, Y.~Suzuki, and K.~Varga, Nucl. Phys. A
\textbf{673}, 122 (2000).\vspace*{0.5mm}

\bibitem {R5}B.E.~Grinyuk and I.V.~Simenog, Yad. Fiz. \textbf{77}, 443 (2014).\vspace*{0.5mm}

\bibitem {R6}A.V. Nesterov, F. Ariks, Ya. Brukkhov, and V.S.~Va\-si\-levs\-kii,
Elem. Chast. At. Yadro \textbf{41}, 1337 (2010).\vspace*{0.5mm}

\bibitem {R7}V.I.~Kukulin and V.M.~Krasnopol'sky, J. Phys. G \textbf{3}, 795 (1977).\vspace*{0.5mm}

\bibitem {R8}N.N. Kolesnikov and V.I. Tarasov, Yad. Fiz. \textbf{35}, 609 (1982).\vspace*{0.5mm}

\bibitem {R9}Y.~Varga and K.~Suzuki, \textit{Stochastic Variational Approach
to Quantum-Mechanical Few-Body Problems} (Springer, Berlin,
1998).\vspace*{0.5mm}

\bibitem {R10}B.E.~Grinyuk and I.V.~Simenog, Ukr. J. Phys. \textbf{56}, 635 (2011).\vspace*{0.5mm}

\bibitem {R11}B.E.~Grinyuk, D.V.~Piatnytskyi, and I.V.~Simenog, Ukr. J. Phys.
\textbf{52}, 424 (2007).\vspace*{0.5mm}

\bibitem {R12}I.~Angeli and K.P.~Marinova, At. Data Nucl. Data Tables
\textbf{99}, 69 (2013).\vspace*{0.5mm}

\bibitem {R13}I.~Sick, Phys. Rev. C \textbf{77}, 041302 (2008).

\bibitem {R14}\textit{Review of Particle Physics}, J. Phys. G \textbf{33},
1--1232 (2006).\vspace*{0.5mm}

\bibitem {R15}I.V. Simenog, M.V. Kuzmenko, and V.M. Khryapa, Ukr. Fiz. Zh.
\textbf{55}, 1240 (2010).\vspace*{0.5mm}

\bibitem {R16}A.I.~Akhiezer and V.B.~Berestetskii, \textit{Quantum
Electrodynamics} (Interscience, New York, 1965).\vspace*{0.5mm}

\bibitem {R17}L.A.~Schaller, L.~Schellenberg, T.Q.~Phan \textit{et al.}, Nucl.
Phys. \textbf{A 379}, 523 (1982).\vspace*{0.5mm}

\bibitem {R18}R.F.~Frosch, J.S.~McCarthy, R.E.~Rand, and M.R.~Yearian, Phys.
Rev. \textbf{160}, 874 (1967).\vspace*{0.5mm}

\bibitem {R19}P.E.~Bosted \textit{et al.}, Phys. Rev. Lett. \textbf{68}, 3841 (1992).\vspace*{3.5mm}

\begin{flushright}
{\footnotesize Received 17.03.16.\\ Translated from Ukrainian by
O.I.~Voitenko}
\end{flushright}
\end{thebibliography}
\end{document}